\documentclass[%
 reprint,
 amsmath,amssymb,
 aps,
]{revtex4-2}

\usepackage{graphicx}% Include figure files
\usepackage{dcolumn}% Align table columns on decimal point
\usepackage{bm}% bold math
\usepackage{float}
\begin{document}

\preprint{APS/123-QED}

\title{Machine Learning for Detecting Steering in Qutrit-Pair States}% Force line breaks with \\

\author{Pu Wang}
\affiliation{	
	School of Control and Computer Engineering, North China Electric Power University, Beijing 102206, China\textbackslash\textbackslash}%Lines break automatically or can be forced with \\
\author{Zhongyan Li}
 %\homepage{http://www.Second.institution.edu/~Zhongyan Li}
%\affiliation{
\author{Huixian Meng}
 \email{huixianmenghd@ncepu.edu.cn}
\affiliation{%
School of Mathematics and Physics, North China Electric Power University, Beijing 102206, China\textbackslash\textbackslash
}%


\date{\today}% It is always \today, today,

\begin{abstract}
Only a few states in high-dimensional systems can be identified as (un)steerable using existing theoretical or experimental methods. We utilize semidefinite programming (SDP) to construct a dataset for steerability detection in qutrit-qutrit systems. For the full-information feature $F_1$, artificial neural networks achieve high classification accuracy and generalization, and preform better than the support vector machine. As feature engineering playing a pivotal role, we introduce a steering ellipsoid-like feature $F_2$, which significantly enhances the performance of each of our models. Given the SDP method provides only a sufficient condition for steerability detection, we establish the first rigorously constructed, accurately labeled dataset based on theoretical foundations. This dataset enables models to exhibit outstanding accuracy and generalization capabilities, independent of the choice of features. As applications, we investigate the steerability boundaries of isotropic states and partially entangled states, and find new steerable states. This work not only advances the application of machine learning for probing quantum steerability in high-dimensional systems but also deepens the theoretical understanding of quantum steerability itself. 
%\begin{description}
%\end{description}
\end{abstract}

\maketitle

%\tableofcontents

\section{\label{section1} Introduction}

Quantum steering was introduced by Schr$\rm{\ddot{o}}$dinger in 1935 to answer the Einstein-Podolsky-Rosen (EPR) paradox \cite{ref1}. It can be used for quantum key distribution \cite{ref2,ref3,ref4}, quantum random number verification \cite{ref5,ref6}, quantum channel resolution \cite{ref7,ref8}, quantum secret sharing \cite{ref9,ref10}, quantum teleportation \cite{ref11,ref12}, etc. To advance the field of quantum information science, a deeper understanding of quantum steering is urgently needed. In this process, Effective detection of quantum state steerability plays a core and fundamental role. Various quantum steering criteria and inequalities have been derived, such as linear quantum steering inequality \cite{ref13,ref14,ref15}, multiplication-based variance inequality \cite{ref16,ref17}, entropy uncertainty relation
\cite{ref18,ref19}, and so on. With the rapid development of computer power, it is now also possible to use semidefinite programming (SDP) \cite{ref20} to determine whether a given quantum state is steering or not through numerical simulations \cite{ref21}.

Recently, machine learning has garnered significant interest for its integration with quantum information, particularly following its successful applications in entanglement discrimination \cite{ref22,ref22+1}, non-locality detection \cite{ref23} and quantum steering detection in two-qubit systems. In 2019, Ren et al. \cite{ref24} first applied the support vector machine (SVM) algorithm for quantum steering of arbitrary two-qubit states. In 2020, Zhang et al. \cite{ref25} used a back-propagation neural network to train quantum steering classifiers, and Zhang et al. (2021) \cite{ref26} applied semi-supervised learning methods to this task. More recently, in 2024, Zhang et al. \cite{ref27} explored steerability detection in qubit-qutrit systems using various machine learning techniques. However, machine learning for high-dimensional quantum steering detection remains limited. High-dimensional quantum steering \cite{ref27+2} has gained attention for its enhanced noise resilience, but there are few related works in both theoretical predictions and experimental confirmations \cite{ref27+3}. In this paper, we extend quantum steering detection to the qutrit-qutrit system, applying machine learning methods to address this challenge.

In this paper, we will detect quantum steering of qutrit-qutrit states via different machine learning techniques. In Sec.\ref{sec:2}, we introduce basic concepts and SDP detecting method of quantum steering. In Sec.\ref{sec:3}, we pioneer the construction of SDP-labeled and accurately labeled datasets for qutrit-qutrit systems. Compared with the full information feature $F_1$, our steering ellipsoid-like feature $F_2$ can train SVM models with higher classification accuracy and more robust generalisation. In the same sense, accurately labeled dataset performs better than SDP-labeled dataset. In Sec.\ref{sec:6} and Sec.\ref{sec:9}, we train ANN and ensemble learning classifiers. Though analysis, we find that ANN models achieve better generalization performance than SVM models for the feature $F_1$, and ensemble learning models perform best for the feature $F_2$. In Sec.\ref{sec:7}, we explore the above trained model to investigate the steerability boundaries of isotropic states and partially entangled states, and find new steerable states. Finally, we summarize our results in Sec.\ref{sec:8}.

\section{\label{sec:2}PRELIMINARIES}

For a bipartite state $\rho^{AB}$ shared by Alice and Bob, Alice performs $m_A$ measurements $\{M_x=\{M_{a|x}\}\}^{m_A}_{x=1}$, where each measurement output $o_A$ corresponds to $a=1,2,\dots,o_A.$ For arbitrary $ m_A\in\mathbb{Z}^+$, if there always exists a Local-Hidden-State (LHS) model: A probability density function $p(\lambda)$ on the hidden variable $\lambda$, Alice's local response functions $\{p(a\mid x,\lambda)\}$, and quantum states $\{\sigma_\lambda\}$ associated with Bob's system satisfy that 
\begin{eqnarray}
	\sigma_{a\mid x}=\int p(\lambda)p(a\mid x,\lambda)\sigma_\lambda d\lambda,\ \forall a,x,
\end{eqnarray}  
where $\sigma_{a\mid x}=tr_A[(M_{a\mid x}\otimes I_B)\rho^{AB}]$, then $\rho^{AB}$ is called {\it unsteerable from Alice to Bob}. Otherwise $\rho^{AB}$ is called steerable from Alice to Bob.

The assemblage $\{\sigma_{a\mid x}\}$ is called unsteerable from Alice to Bob if there exists a LHS model such that (1) holds. Otherwise we classify $\{\sigma_{a\mid x}\}$ as steerable from Alice to Bob. It is difficult  that  runs over comprehensive measurements to determine whether a quantum state is unsteerable. However the unsteerability of the assemblage $\{\sigma_{a\mid x}\}$ can be detected by SDP. Let’s quickly revisit the SDP method \cite{ref20}. 

Suppose that Alice performs measurements $\{M_x=\{M_{a|x}\}\}^{m_A}_{x=1}$. Then the assemblage $\{\sigma_{a\mid x}\}$ is unsteerable from Alice to Bob if the following semi-definite program
\begin{eqnarray}
	&&given\ \ \ \   \{\sigma_{a\mid x}\},\{D(a\mid x,\lambda)\} \nonumber\\
	&&find\  \ \ \ \ \ \{\sigma_\lambda\},\ \ \  \sigma_\lambda \ge0 \\
	\displaystyle  &&s.t.\ \ \ \ \ \  \sum_{\lambda}D(a\mid x,\lambda)\sigma_\lambda=\sigma_{a\mid x}\nonumber
\end{eqnarray}
is solvable, where $\{D(a\mid x,\lambda)\}$ is the deterministic single-party conditional probability distribution, which gives a fixed outcome $a$ for each measurement $x$, that is, $D(a|x,\lambda)=1$ if $\lambda(x)=a$ and $D(a|x,\lambda)=0$ if $\lambda(x)\ne a$.
The dual problem of (2) is:
\begin{eqnarray}
	&&given\ \   \{\sigma_{a\mid x}\},\{D(a\mid x,\lambda)\} \nonumber\\
	&&{min}_{F_{a\mid x}} \ \ tr(\sum_{a,x}F_{a\mid x}\sigma_{a\mid x})\\
	&&s.t.\ \ \sum_{a,x}D(a\mid x;\lambda)F_{a\mid x}\geq 0 \nonumber
\end{eqnarray}	
where $F_{a\mid x}$ are Hermitian matrices. If the optimal objective value is negative for some measurement $x$, then $\rho^{AB}$ is steerable from Alice to Bob. On the other hand, a non-negative value means that assemblage $\{\sigma_{a\mid x}\}$ is unsteerable.

To assess the steerability of a given quantum state, the SDP method demands extensive measurements, which is time-consuming in practice. To improve detection efficiency of two-qubit states, machine learning has been applied in Ref.\cite{ref24,ref25,ref26}.To tackle challenges in detecting steerability of two-qutrit states, we generated two types of datasets—SDP-labeled and accurately labeled—and employed three machine learning models: SVM, ANN, and ensemble learning algorithms, along with two feature sets, $F_1$ and $F_2$ ($F'_1$ and $F'_2$). To avoid any confusion, the following systems will be referred to as qutrit-qutrit system, unless otherwise specified, and this paper focuses only on the states are (un)steerable Alice to Bob.

\section{\label{sec:3} Detecting the steerability by SVM}
Theoretically, as the number of SDP measurements increases, quantum steering detection precision improves. However, in high-dimensional quantum systems, the SDP approach is notably time- and space-consuming to ensure some extent of accuracy because of the trade-off between accuracy and computational efficiency. To address these challenges, we integrate SDP with supervised learning, as machine learning has the power to accelerate the computation process.

SVM is a supervised machine learning method that aims to find the best hyperplane that separates different samples as much as possible \cite{ref28}, which requires solving the following optimization problem:
\begin{eqnarray}
	&&{min}_{\omega,b,\xi_i}\ \ \  \frac{1}{2}\Arrowvert\omega\Arrowvert^2+C\sum_{i=1}^l\xi_i\nonumber	\\
	&&s.t.\ y_i(\omega^T\phi(\bm{x_i})+b)\geq1-\xi_i\\
	&&\ \ \ \ \xi_i\geq0\ \ \ i=1,2,\cdots,l.\nonumber
\end{eqnarray}
where $l$ is the number of samples, $\bm{x_i}$ $,y_i $ denote the sample vectors and labels respectively, and $\phi$ is the mapping determined by the kernel function. 

In order to train the quantum steering classifiers, we need to collect the data for quantum states and select the features of the data.

\subsection{\label{sec:4}Data production with feature $F_1$}
\subsubsection{Data production}
Any qutrit-qutrit state can be described by a density matrix of $9 \times 9$. Analogous to the generation method of random state datasets in qubit-qubit system, we can generate random state datasets in qutrit-qutrit system through the following steps:

\begin{itemize}
	\item A density matrix $\rho_{AB}:=H/Tr(H)$ is generated by two random $9 \times 9$ matrices $M$ and $N$, where $H := (M+iN)(M+iN)^\dag$ with $\dag$ being the conjugate transpose. 
	
	\item Extract full information feature $F_1$: The initial eight components on the primary diagonal and the real and imaginary parts of 36 elements below the diagonal of the density matrix $\rho_{AB}$ are extracted to constitute an 80-dimensional vector, which is one-to-one correspondence with the density matrix. We denote this vector as feature $F_1$.
	
	\item Alice randomly generates $m$ measurements $\hat{M_k}=\vec{n_k}\cdot\hat{S},$ with $\vec{n_k}=(sin\theta_kcos\phi_k,sin\theta_ksin\phi_k,cos\theta_k)$ and $\hat{S}=(S_x,S_y,S_z)$. The spin-1 operators 
	{\small
		\begin{eqnarray} S_x &=& \frac{1}{\sqrt{2}}\left(
			\begin{array}{ccc}
				0 & 1 & 0 \\
				1 & 0 & 1 \\
				0 & 1 & 0 \\
			\end{array}
			\right),
			S_y=\frac{1}{\sqrt{2}}\left(
			\begin{array}{ccc}
				0 & -i & 0 \\
				i & 0 & -i \\
				0 & i & 0 \\
			\end{array}
			\right),\\
			S_z &=& \frac{1}{\sqrt{2}}\left(
			\begin{array}{ccc}
				1 & 0 & 0 \\
				0 & 0 & 0 \\
				0 & 0 & -1 \\
			\end{array}
			\right)\nonumber\end{eqnarray}}introduced in \cite{ref27+1} can be used to express the basis of observables on qutrit system in the sense that any Gell-Mann matrix is a linear combination of $S_x,\ S_y,\ S_z,\ S^2_x,\ S^2_y,\ \{S_x,S_y\},\ \{S_y,S_z\}$ and $\{S_z,S_x\}$, where $\{S_i,S_j\}=S_iS_j+S_jS_i(i,j=x,y,z)$ denotes the corresponding anticommutator. To simplify computation, Alice chooses measurements with the form $\hat{M_k}=\vec{n_k}\cdot\hat{S}$. Since eigenvalues of $\hat{M_k}$ are $-1,0,1$, we use $P_{-1|x},P_{0|x},P_{1|x}$ to denote the corresponding eigenprojectors.
	
	\item For any $k$, one hundred pairs of $(\theta_k, \phi_k)$ are randomly sampled to produce 100 assemblages, each of which is individually inspected utilizing SDP. If a negative value has been obtained, then $\rho^{AB}$ is steerable from Alice to Bob. In this case, the corresponding feature is labeled as $-1$. Otherwise, this feature the label $+1$, which means that we do not know whether $\rho^{AB}$ is steerable.
\end{itemize}

For each $m = 3,\cdots,7$, we generate the corresponding dataset until at least 4000 examples with the label $+1$ and 4000 examples with the label $-1$ are obtained.

\subsubsection{Training and testing}
SVM models are trained using feature $F_1$, which encodes the full information of a quantum state. Firstly, we opt for a soft-margin SVM with a Gaussian kernel function $K(\phi(\bm{x_i}),\phi(\bm{x_j}))=e^{-\gamma\Arrowvert x_i-x_j\Arrowvert^2}$. Then we train the SVM parameters C and $\gamma$ in (4) with the methods of five-fold cross-validation and grid search. In this process, the random state dataset is divided into 6 parts: 5 parts as the training set and 1 part as the test set. To obtain the accuracy of SVM prediction, we use the well-trained SVM to detect the steering of random state in the test set.

The Fig.~\ref{Fig.1} shows that the classification accuracy of SVM models on random states can be kept over $85\%$ (See appendix \ref{tab1} for detailed data) except $m=4$. In the case $m=4$, the classification accuracy can be improved to 89.4\% on train set and the test accuracy improved to 87.5\% by regenerating 20000 random states with 10000 positive and 10000 negative samples. Consistent with our intuition, increasing the amount of data can improve the classification accuracy of the model.

\begin{figure}[h]
	\centering
	\includegraphics[width=8.5cm]{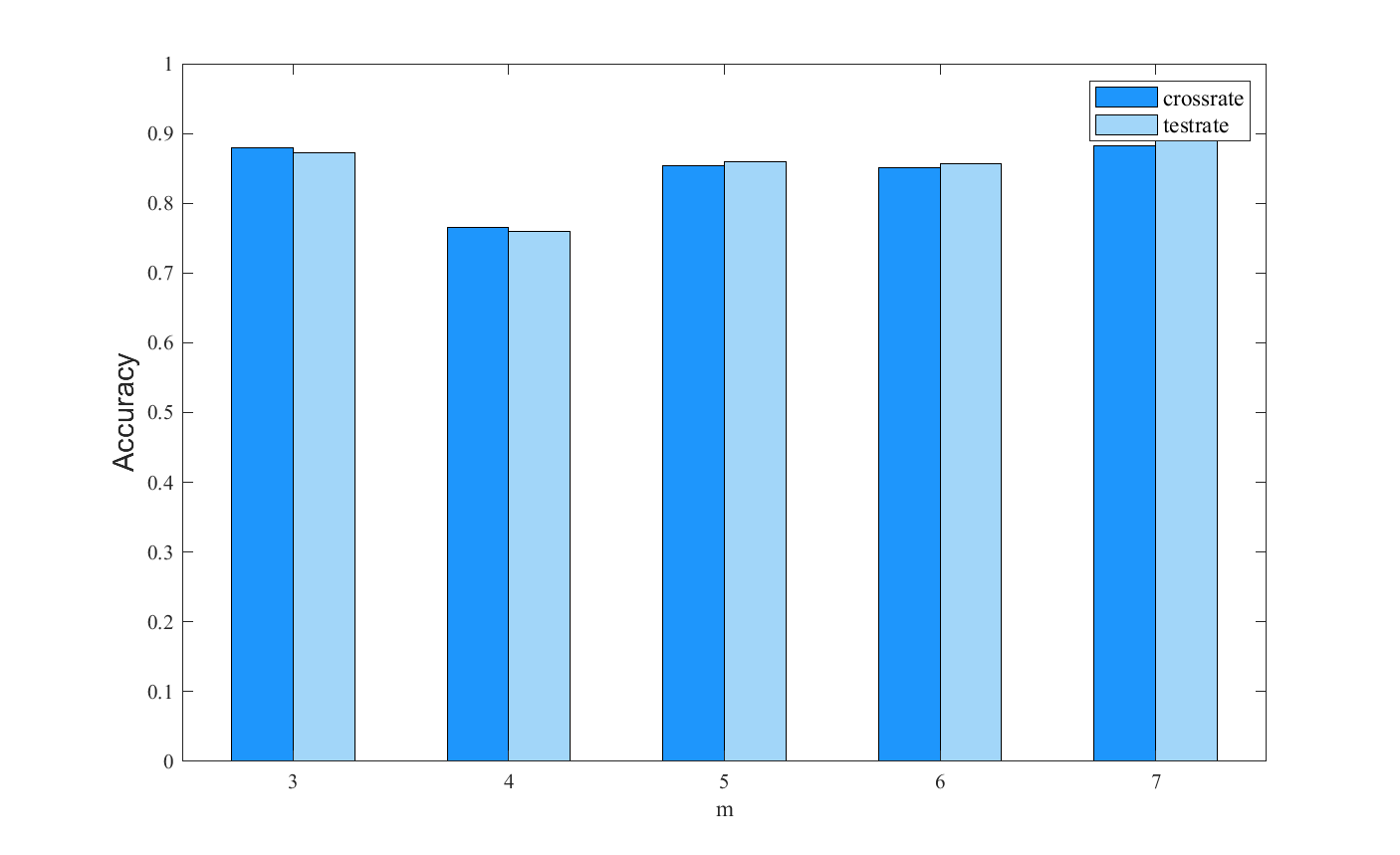}
	\caption{Classification accuracy of SVM. Dark blue indicates cross-validation accuracy and light blue indicates classification accuracy on the test set.} 
	\label{Fig.1}
	%\ref{fig:example1}
\end{figure}

To assess the generalization ability of these SVM classifiers, we examine their performance in classifying a specific class of quantum states with a precise steering bound. In this paper, we chose
this class of quantum states as isotropic states,
\begin{eqnarray}
	S^\eta_3=\eta|\psi_+\rangle\langle\psi_+|+(1-\eta)\frac{I_{9}}{9}, 
\end{eqnarray}
where $|\psi_+\rangle=\frac{1}{\sqrt{3}}\sum_{i=1}^{3}|i,i\rangle.$ Let $H_3=\sum_{i=1}^{3}\frac{1}{i}.$ Then $S_3^\eta$ is steerable if and only if $\eta>\frac{H_3-1}{2}.$

\begin{figure}[h]
	\centering 
	\includegraphics[width=8.5cm]{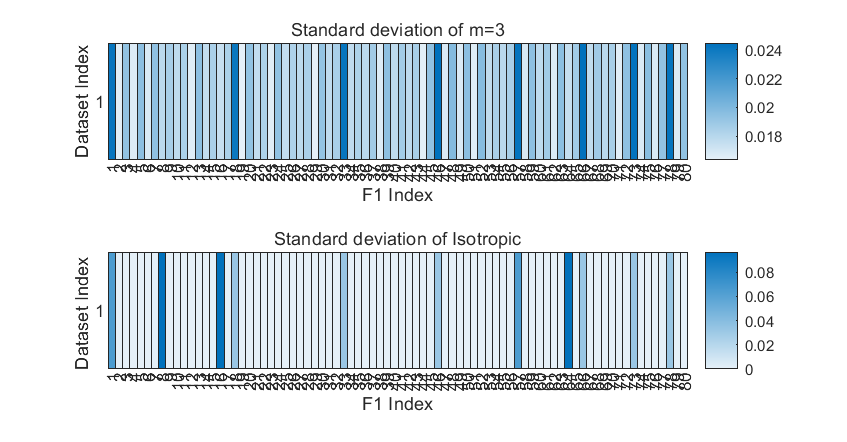}
	\caption{The standard deviation heatmaps of dataset. The first picture is the standard deviation heatmaps of the training set data characterized by $F_1$ for the random state $(m=3)$. The second picture is the standard deviation heatmaps of the test set data characterized by $F_1$ for the isotropic states.} 
	\label{Fig.2}
\end{figure}

We uniformly select the parameter $\eta$ within intervals $[0, \frac{H_3-1}{2}]$ and $[\frac{H_3-1}{2},1]$ to generate 2000 examples of unsteering isotropic states and 2000 examples of steering isotropic states, respectively. Then we apply the above pre-trained SVM models to classify the dataset of isotropic states. However, the classification accuracy remains at $50\%$, indicating a lack of robust generalization performance.
Compared with the standard deviation heatmaps of the training set data,  characterized by $F_1$ for the random state $(m=3)$, the corresponding heatmaps for isotropic states has more locations containing zeros. See Fig. \ref{Fig.2}.  The reason for the lack of generalisation ability of our SVM models may be that the test set generated by isotropic states is in the margin of random states set.

\subsection{Data with feature $F_2$ production}

Features significantly impact how a model recognizes objects. In the reference \cite{ref29}, it is noted that for two-qubit quantum systems, the most compact feature of Alice-to-Bob steerability is provided by Alice's regularly aligned quantum steering ellipsoid. We generalize this feature to qutrit-qutrit systems, deriving a steering ellipsoid-like feature $F_2$.

\subsubsection{Data production}
Arbitrary 2-qutrit state can be written as
$$\rho^{AB}=\frac{1}{9}I_9 + \frac{1}{6}(\vec{a}\cdot\vec{\delta}\otimes I_3 +I_3\otimes\vec{b}\cdot\vec{\delta})+\frac{1}{4}\sum_{i,j=1}^{8}T_{ij}\delta_i\otimes\delta_j,$$ 
in the Gell-Mann basis \cite{ref30}. When we take $\Phi_{ij}=Tr(\rho^{AB}\delta_i\otimes\delta_j)(\forall i,j\in\{0,1,\dots,8\})$, $\Phi=(\Phi_{ij})$ has a block structure:
$$\Phi=\left(
\begin{array}{cc}
	1 & \vec{b}^T\\
	\vec{a} & T\\
\end{array}
\right).
$$
If $(\rho^B)^{\frac{1}{2}}$ is invertible, then we can obtain a new $\tilde{\rho}^{AB}$, which has the same steerability with $\rho^{AB}$, by an one-way stochastic local operations and classical communication (1W-SLOCC):
$$\rho^{AB}\mapsto\tilde{\rho}^{AB}=\frac{(I_3\otimes(\rho^B)^{\frac{-1}{2}})\rho^{AB}(I_3\otimes(\rho^B)^{\frac{-1}{2}})}{tr\left((I_3\otimes(\rho^B)^{\frac{-1}{2}})\rho^{AB}(I_3\otimes(\rho^B)^{\frac{-1}{2}})\right)}.$$
The matrix $\tilde{\Phi}$ corresponding to $\tilde{\rho}^{AB}$ is
$$\tilde{\Phi}=\left(
\begin{array}{cc}
	1 & \vec{0}^T\\
	\vec{\tilde{a}} & \tilde{T}\\
\end{array}
\right),
$$
where $\tilde{T}$ is an $8\times 8$ real matrix and $\tilde{T_{ij}}=Tr(\tilde{\rho}^{AB}\delta_i\otimes\delta_j)(\forall i,j\in\{1,2,\dots,8\}).$
We can diagonalize $\tilde{T}$ by singular value decomposition
$$\tilde{T}=O_1T'O^T_2,$$ 
where $T'=(diag(\lambda_1),\dots,diag(\lambda_8))$ and $\lambda_i(i=1,\dots,8)$ are singular value of $\tilde{T}$. Then we can find an appropriate local unitary transformation by $$U_X\vec{n}\cdot \vec{\delta}U_X^\dagger=(O_i\vec{n})\cdot\vec{\delta}\ (X=A,B,i=1,2)$$ such that
$$\tilde{\rho}^{AB}\mapsto\rho'^{AB}=(U_A\otimes U_B)\tilde{\rho}^{AB}(U_A\otimes U_B)^\dagger.$$

The corresponding matrix can be obtained
$\Phi'$:
$$\Phi'=\left(
\begin{array}{cc}
	1 & 0^T\\
	\vec{a'} & T'\\
\end{array}
\right).
$$

Since the most compact feature of Alice-to-Bob steerability is provided by Alice's regularly aligned quantum steering ellipsoid \cite{ref31,ref32,ref32+1}, we extract diagonal elements of $Q'_A=T'T'^T$ and $\vec{a'}=O_1\vec{\tilde{a}}$ as features $F_2$ with length of 16. 

We convert the existing datasets with feature $F_1$ into datasets with feature $F_2$, which is composed as follows: for each $m=3,5,6,7,$ we generate the corresponding datasets with 4000 examples with the label $-1$ and 4000 examples with the label $+1$; for $m=4$, we can get the dataset with 10000 examples with the label $-1$ and 10000 examples with the label $+1$.

\subsubsection{Training and testing}
Similar to the research in Sec.3.1.2, we can train SVM models for feature $F_2$, and obtain the accuracy of these well-trained SVM models in the test set. It is encouraging that they not only slightly improve the classification accuracy on random states, but also have robust generalization performance, i.e., they can classify the dataset of isotropic states with the accuracy over $89.7\%$ (The details of the data are in Tab.~\ref{tab2}). 

\begin{figure}[h]
	\centering
	\includegraphics[width=8.5cm]{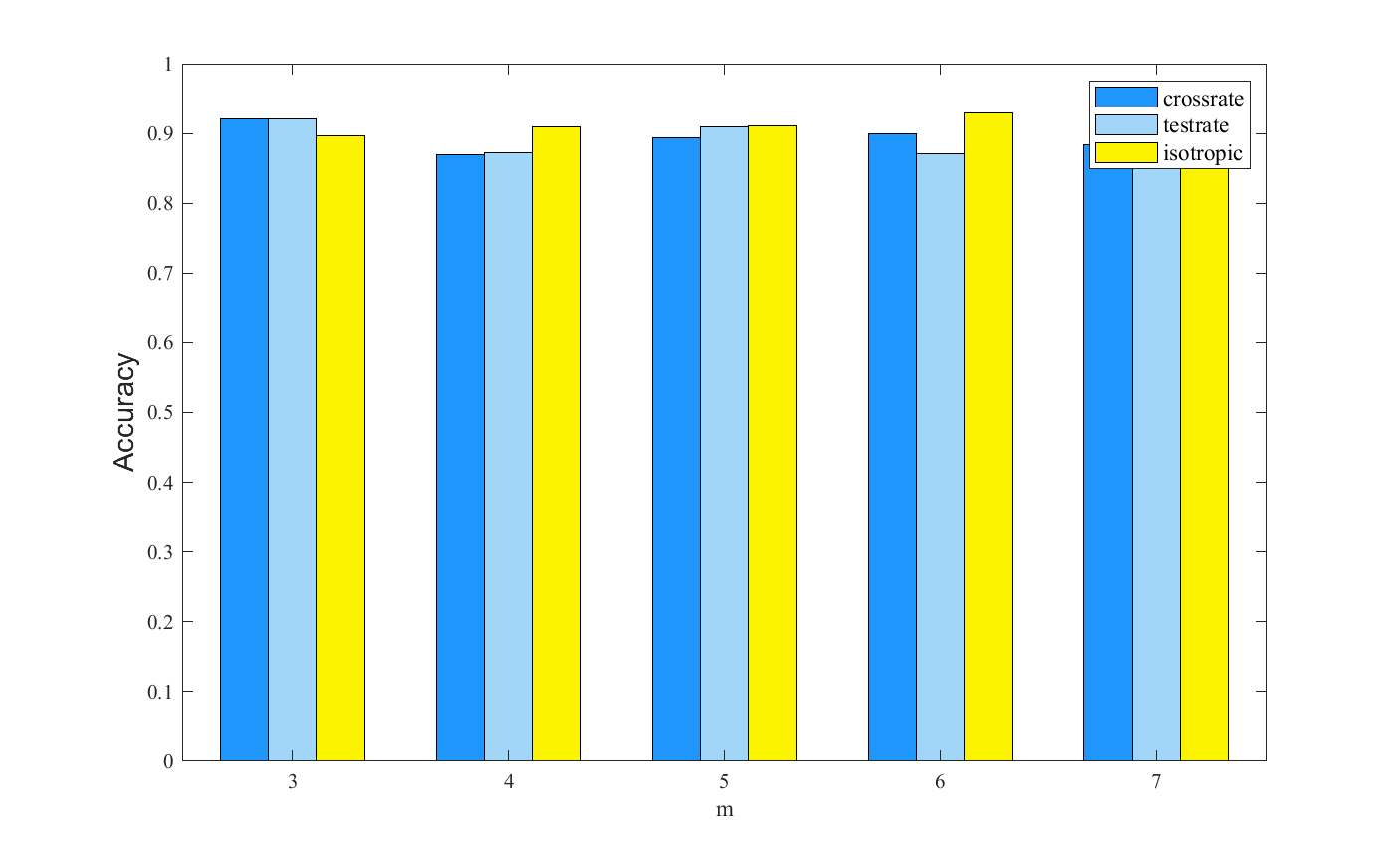}
	\caption{Classification accuracy of SVM. Dark blue indicates cross-validation accuracy and light blue indicates classification accuracy on the random states test set. Yellow indicates classification accuracy on the isotropic states dataset} 
	\label{Fig.3}
	%\ref{fig:example1}
\end{figure}

In Fig. \ref{Fig.3}, we plot the accuracy of SVM models' prediction with feature $F_2$. It indicates that feature $F_2$ is significant for training robust SVM classifiers.

For general random states, the unsteerability can be detected only when the SDP method runs over all measurements. It is non-implementable in principle. This highlights the preciousness of datasets with accurate labels. In the following, we construct the dataset by SDP method and relevant theoretical results.

\subsection{Data production with accurate labels}
\subsubsection{Data production}

To produce the datasets with accurate labels, we collect steerable quantum states, including entangled pure states, steerable isotropic states, steerable Werner states, and $m \le 7$ random states identified as steerable by the SDP, as the data labeled $-1$; and we collect unsteerable quantum states, including separable pure states, separable mixed states, unsteerable Werner states and unsteerable isotropic states, as the data labeled $+1$.

For each class of quantum states, we randomly generated 2000 data with features $F'_1$ and $F'_2$, respectively (The features $F'_1$ and $F'_2$ are extracted in the same manner as $F_1$ and $F_2$ described earlier. For clarity and distinction, we refer to them as $F'_1$ and $F'_2$).

Since the Werner states and isotropic states are in the training dataset, we chose partially entangled states to test the generalisation ability of our trained machine learning models. The states can be written as
$$\rho(p,\theta,\phi)=p|\psi_{\theta,\phi}^+\rangle\langle\psi_{\theta,\phi}^+|+(1-p)\rho^A_{\theta,\phi}\otimes I/3,$$
where $|\psi_{\theta,\phi}^+\rangle=\cos(\theta)\sin(\phi)|00\rangle+\sin(\theta)\sin(\phi)|11\rangle+\cos(\phi)|22\rangle,$ $p\in[0,1]$, $\theta\in[0,\pi/4]$, and $\phi\in[0,\pi/2]$. With SDP method, reference \cite{ref33} found partially entangled states with parameter $p<0.4818$ were unsteering by performing four-setting $(m=4)$ mutually unbiased bases measurements.
We uniformly chose the parameter $p$ within the intervals $[0, 1]$, the parameter $\theta$ within the interval $[0,\pi/4]$, and the parameter $\phi$ within the interval $[0,\pi/2]$ to generate 2000 examples of unsteering partially entangled states and 2000 examples of steering partially entangled states. 
%Further we construct the test set with features $F_1$ and $F_2$.
\subsubsection{Training and testing}
Similar to the research based on feature $F_1$ and $F_2$, we can train SVM models for feature $F'_1$ and $F'_2$, and obtain the accuracy of these well-trained SVM models in the test set.

Based on the results in Tab.~\ref{tab3}, it is evident that the well-trained SVM models with accurately labeled datasets, in contrast to the original datasets, yields significantly better performance. Specifically, the accurately labeled datasets outperform the originals in terms of cross-validation accuracy, testing accuracy, and generalization capability. These findings strongly indicate that the use of accurate labels contributes to reduced quantum steering classification errors and enhances the overall effectiveness of SVM model training.

\begin{table}[h]
	\centering
	\begin{ruledtabular}
		\caption{\label{tab3}the accuracy of SVM prediction with accurate label}
		\begin{tabular}{ccc}
			SVM & $F'_1$ & $F'_2$  \\
			\hline
			crossrate & 99.10\% & 96.60\% \\
			testrate & 96.20\% & 97.20\% \\
	\end{tabular}\end{ruledtabular}
\end{table}

\section{\label{sec:6}Detecting the steerability by the artificial neural network}
\begin{figure}[h]
	\centering
	\includegraphics[width=8cm]{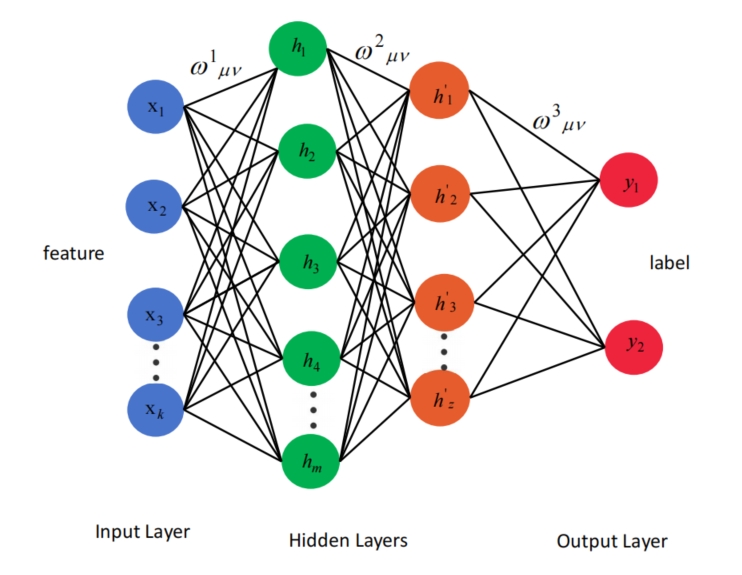}
	\caption{\label{Fig.4}The structure of an ANN.} 
	%\ref{fig:example1}
\end{figure}

 The ANN is one of the most widely used deep learning models \cite{ref34}, which can be used to infer the steerability of two-qubit states \cite{ref29}. In this section, we leverage ANN models to detect the steerability of two-qutrit states. As shown schematically in Fig. \ref{Fig.4}, an ANN consists of an input layer, several hidden layers, and an output layer. The number of neurons in the input layer is the length $k$ of the feature. In our constructions,  (1) $k=80$ for the feature $F_1(F'_1)$ and $k=16$ for the feature $F_2(F'_2)$; (2) the number of hidden layers is two; (3) the weights $\{w^{(i)}_{uv}\}$ connecting the neurons between layers are optimized with the conventional backpropagation algorithm aiming at minimizing the loss function; (4) the outputs of the neurons in the hidden layers are determined by an activation function ReLu.

\subsection{Training and testing}
For the datasets with feature $F_1$, the two-layer neural network is trained by the five-fold cross-validation method with the number of training $N = 1000$ and the learning rate $w = 0.1$.

\begin{figure}[h]
	\centering 
	\includegraphics[width=8.5cm]{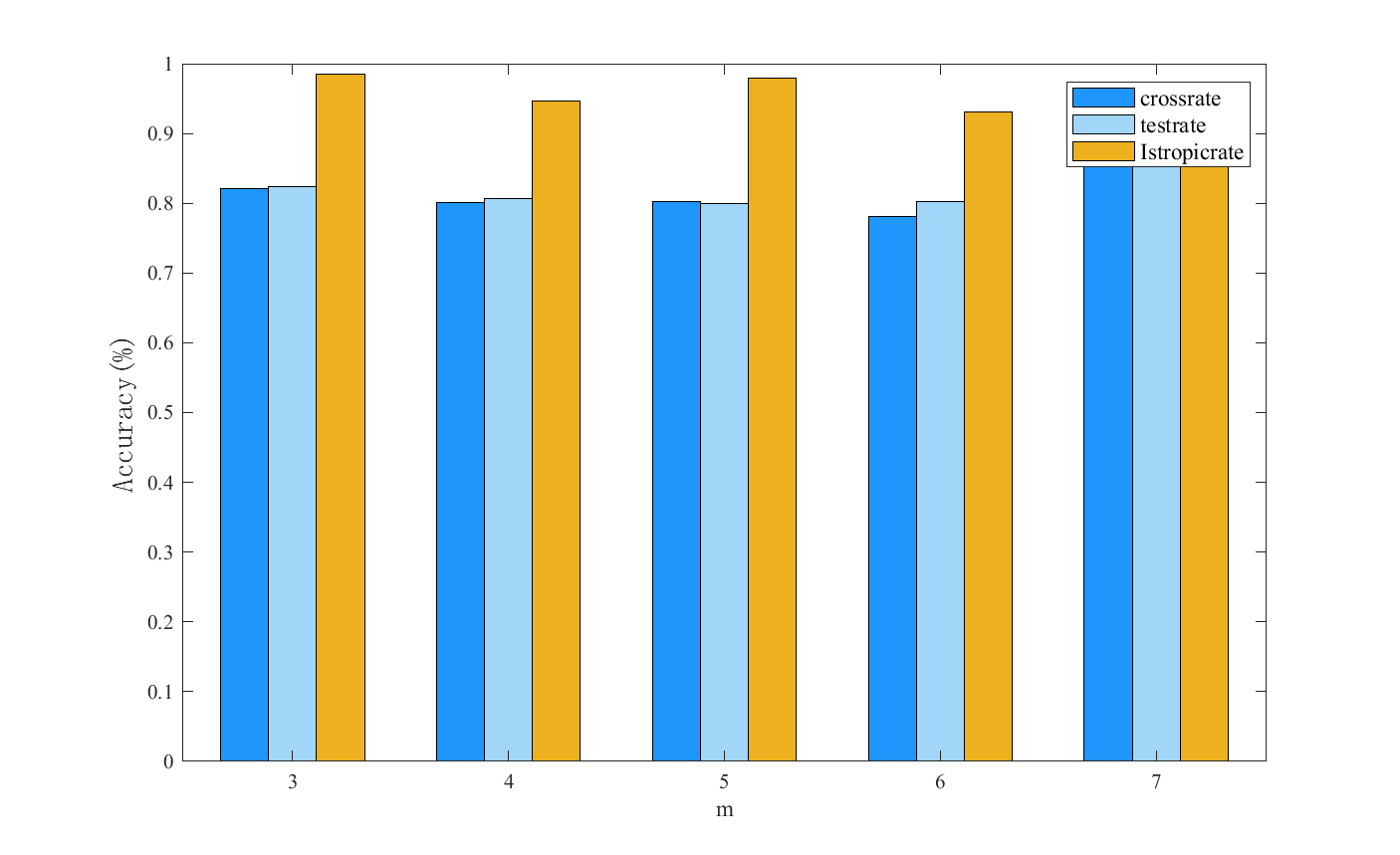}
	\caption{Classification accuracy of ANN. Dark blue indicates cross-validation accuracy and light blue indicates classification accuracy on the random state test set. Yellow indicates classification accuracy on the Isotropic state datasets}
	\label{Fig.5}
\end{figure}

The Fig. \ref{Fig.5} shows that the classification accuracy of ANN models on random states can be around $80\%$ (The details of the data are in Tab. \ref{tab4}).  Despite the fact that the classification accuracy in the Fig. \ref{Fig.5} is slightly lower than that in Fig. \ref{Fig.1}, received by SVM models, however the ANN models have robust generalization ability on test set of isotropic states. 

\begin{figure}[h]
	\centering 
	\includegraphics[width=8.5cm]{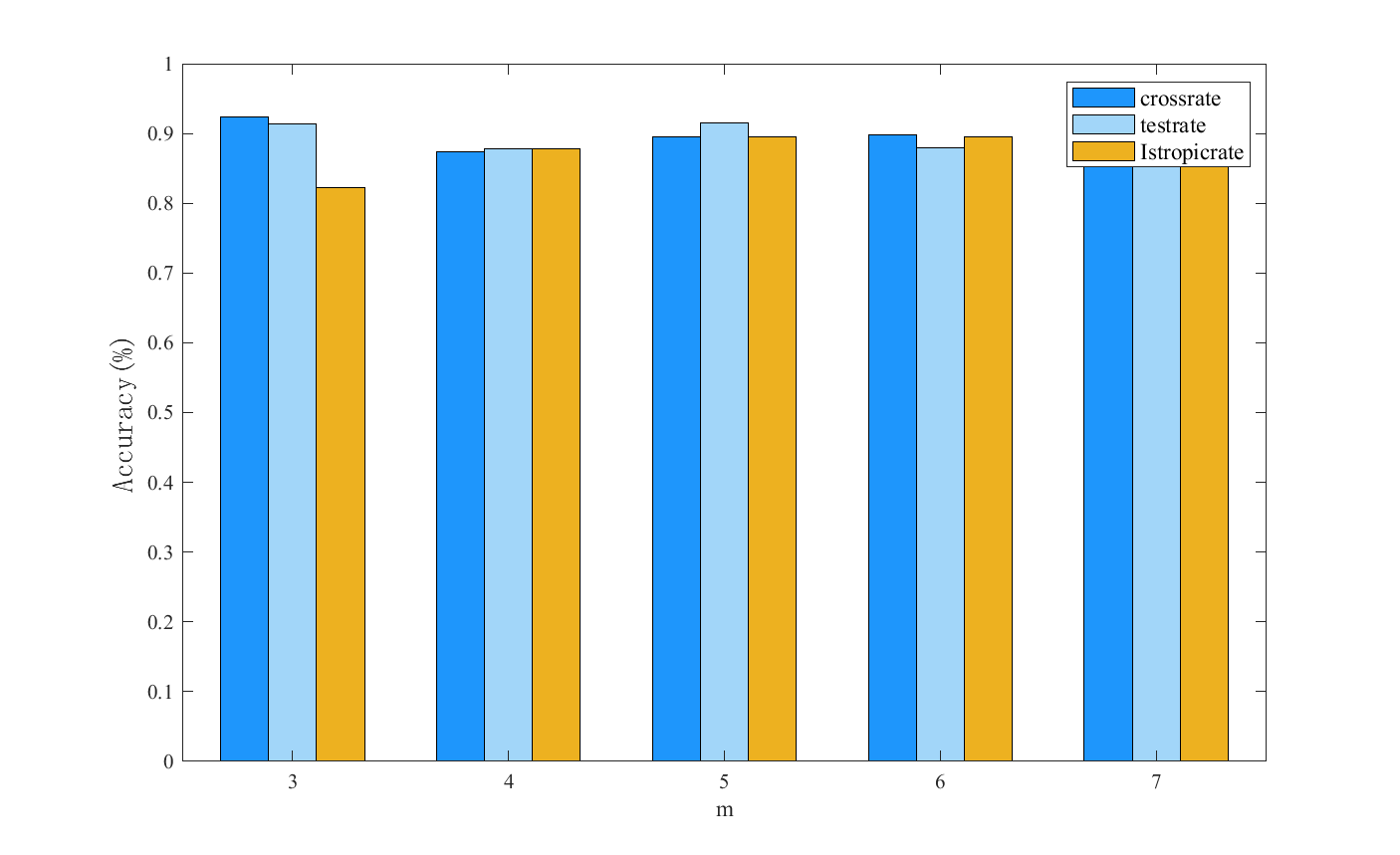}
	\caption{Classification accuracy of ANN. Dark blue indicates cross-validation accuracy and light blue indicates classification accuracy on the random state test set. Yellow indicates classification accuracy on the Isotropic state datasets}
	\label{Fig.6}
\end{figure}

Exchange the feature $F_1$ for the feature $F_2$, the classification accuracy of ANN models is improved on random states. The Fig. \ref{Fig.6} shows that the accuracy of ANN models prediction with feature $F_2$ is improved to around 87\% on random states (The details of the data are in Tab. \ref{tab5}).

The above results show that compared to the SDP-generated datasets with feature $F_1$, ANN models has higher test accuracy on the SDP-generated datasets with feature $F_2$. The ANN models can effectively classify quantum steering for all datasets with SDP labels, and the generalisation ability is superior to the SVM models. 

Whereas the excellent performance of the SVM models on accurately labeled datasets, we use the same high-quality datasets to train and test ANN models. As shown in Tab. \ref{tab6}, the ANN models perform very well on the accurately labeled datasets, exhibiting not only high accuracy but also robust generalization capabilities. Notably, when comparing the datasets with the accurate labels for feature $F'_2$, the ANN models achieve higher cross-validation accuracy, testing accuracy, and overall generalization ability on the datasets with feature $F'_1$.
\begin{table}[h]
	\centering
	\footnotesize
	\caption{the accuracy of ANN prediction with accurate label}\label{tab6}
	\begin{ruledtabular}
		\begin{tabular}{ccc}
			ANN & $F'_1$ & $F'_2$  \\
			\hline
			crossrate & 99.50\% & 97.40\% \\
			testrate & 99.70\% & 98.20\% \\
			partially-entangle & 99.50\% & 94.50\%\\
	\end{tabular}\end{ruledtabular}
\end{table}

\section{\label{sec:9}Detecting the steerability by ensemble learning}

Combine several weak learners into one strong learner, an ensemble learning  model has higher accuracy and stability than individual ones \cite{ref35}. In Ref. \cite{ref27}, XGBoost, a kind of ensemble learning model, was used to infer the steerability of qubit-qutrit states. In this section, we leverage  ensemble learning models consisting of decision trees to detect the steerability of qutrit-qutrit states.

\begin{figure}[htbp]
	\centering 
	\includegraphics[width=8cm]{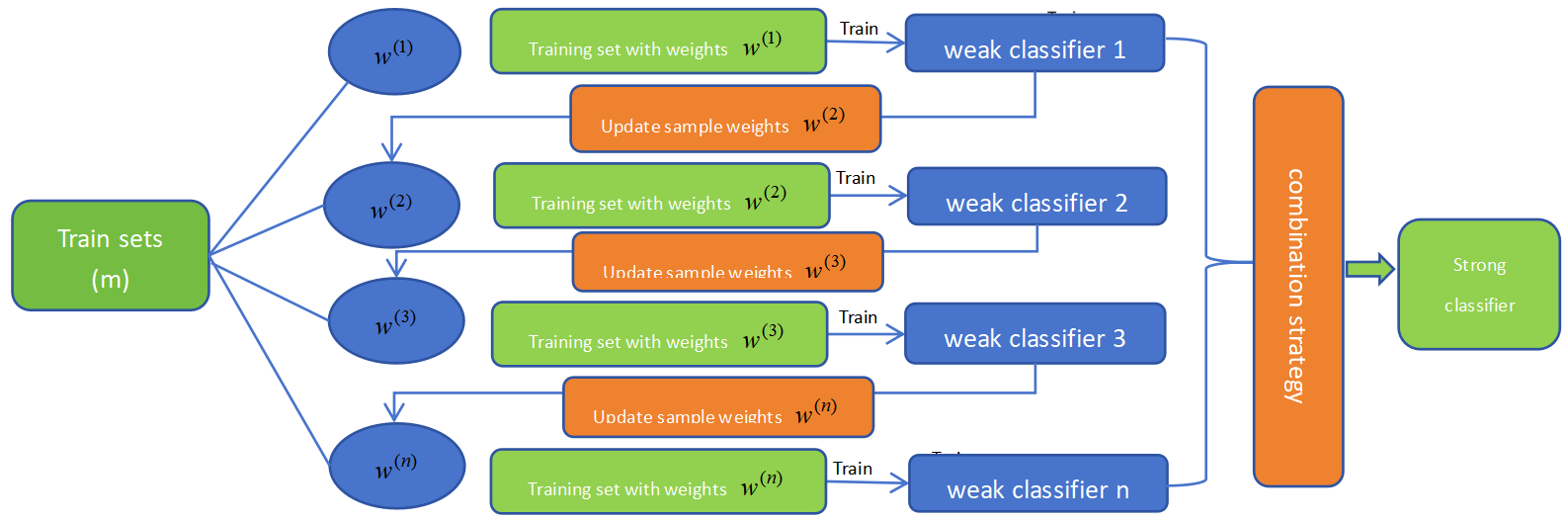}
	\caption{The procedure of a boosting method.}
	\label{Fig.7}
	%\cref{Fig.7}
\end{figure}

Ensemble learning methods can be broadly classified into two main categories: boosting and bagging. Boosting is an algorithm that enhances a weak model into a strong one, and its operational mechanism is as follows: First, train a weak model using the initial training set. Next, the distribution of the training samples will be adjusted based on the performance of this weak model. This means giving more attention to the samples that the weak model misclassifies. Then, train the next weak model using the adjusted samples. Repeat this process until you reach a specified number of weak models or until the performance metrics meet expectations. Finally, these weak models are combined by weighting and summing them to create a strong model. For the convenience of discussion, we presented the procedure of the boosting method in Fig.~\ref{Fig.7}.

\subsection{Training and testing}

\begin{figure}[h]
	\centering 
	\includegraphics[width=8.5cm]{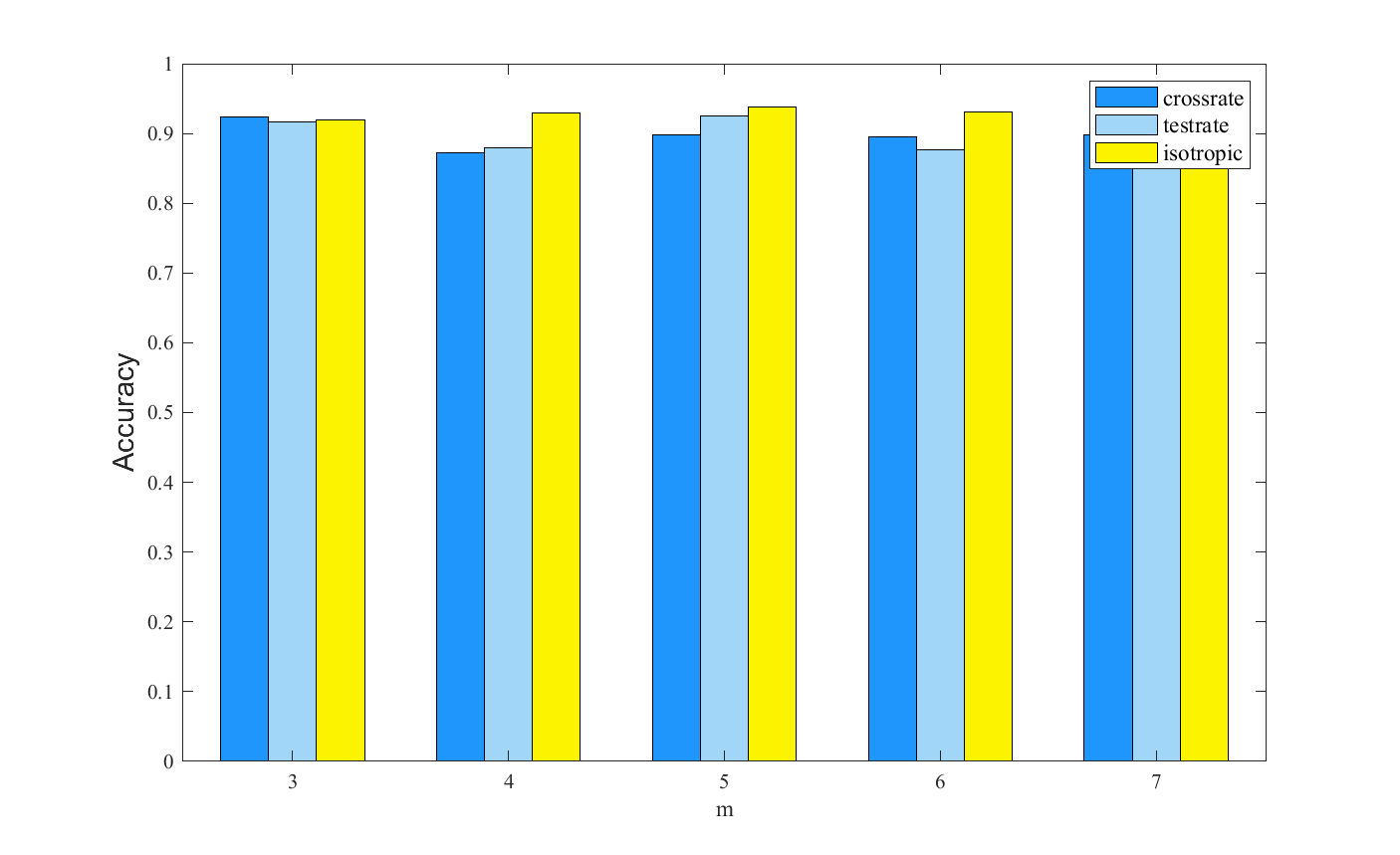}
	\caption{Classification accuracy of ensemble learning. Dark blue indicates cross-validation accuracy and light blue indicates classification accuracy on the random states test set. Yellow indicates classification accuracy on the isotropic states dataset.}
	\label{Fig.8}
	%\cref{Fig.6}
\end{figure}

When we use the datasets with feature $F_1$, no decision tree can classify successfully. Exchange the feature $F_1$ for the feature $F_2$, the classification accuracy of ensemble learning models is the highest compared with SVM and ANN models. The Fig.~\ref{Fig.8} shows that ensemble learning models achieve an accuracy of around 87.5\% for random states steering classification and over 91\% for isotropic states classification (The details of the data are in Tab.~\ref{tab7}).

In light of the excellent performance of the ensemble learning models on datasets with feature $F_2$ and the significance of the datasets with accurate labels, we use the feature $F'_2$ to train and test ensemble learning models. They perform best compared with SVM and ANN models. 

For the accurately labeled dataset with features $F'_1$, the ensemble learning model not only can classify random states with classification accuracy over 97.0\%, but also have robust generalization performance on partially entangled states.

\begin{table}[h]
	\centering
	\footnotesize
	\caption{the accuracy of ensemble learning prediction with accurate label}\label{tab8}
		\begin{ruledtabular}
		\begin{tabular}{ccc}
			Ensemble & $F'_1$ & $F'_2$  \\
			\hline
			crossrate & 97.60\% & 99.40\% \\
			testrate & 97.20\% & 99.80\% \\
			partially-entangled & 96.30\% & 93.80\%\\
	\end{tabular}\end{ruledtabular}
\end{table}

\section{\label{sec:7}Detection of quantum steering bounds}
In this section, we focus on predicting the quantum steering bounds of isotropic states and partially entangled states. For the SDP-labeled datasets, the feature $F_2$ performs better than feature $F_1$, in the sense that SVM, ANN and ensemble learning models exhibit not only high classify accuracy but also robust generalization capabilities. Therefore, we choose the feature $F_2$ to predict quantum steering bounds of isotropic states.

\begin{figure}[h]
	\centering 
	\includegraphics[width=8cm]{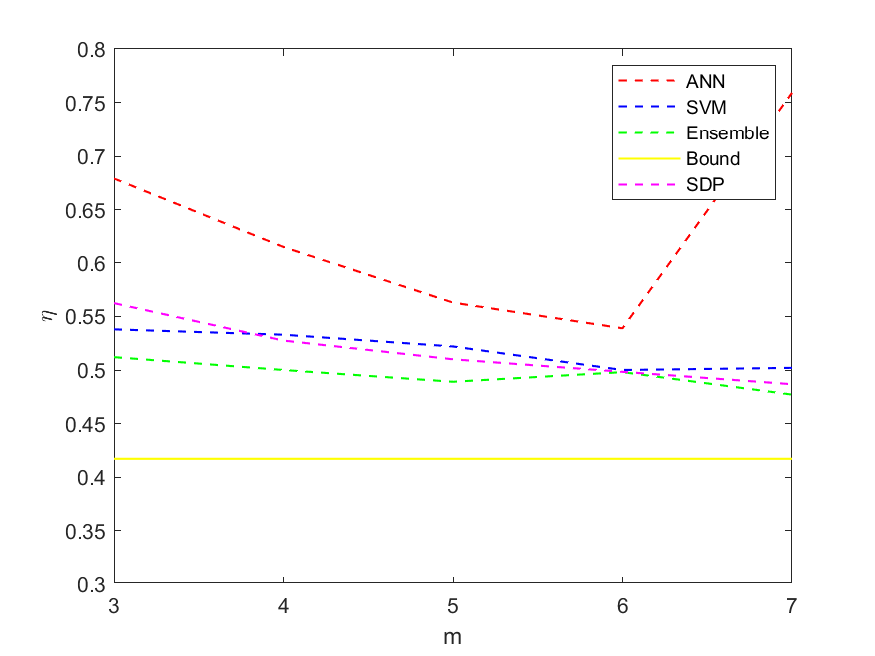}
	\caption{The predictions of steerability for isotropic state by machine learning classifiers. The yellow solid line is the steerability bound from Alice to Bob which is defined by the theory. The dotted blue line is predicted by SVM with $m=3,\dots,7$. The red dotted line is the result predicted by ANN with $m=3,\dots,7$. The green dotted line is the result predicted by ensemble learning models with $m=3,\dots,7$. The magenta dotted line is the result predicted by SDP with $m=3,\dots,7$.}
	\label{Fig.9}
	%\cref{Fig.6}
\end{figure}

Isotropic states $S^{\eta}_3$ in Eq.(5) is unsteerable from Alice to Bob if and only if $\eta \in [0, \frac{H_3-1}{2}]$. In Fig.~\ref{Fig.9}, the yellow solid line $\eta=\frac{H_3-1}{2}$ is plotted as a reference; the steerability of $S^{\eta}_3$ with $\eta$ higher than corresponding dotted line can be detected by the corresponding classifier, i.e., all dotted lines are the steerable bounds predicted by our machine learning models and SDP. Since dotted lines are higher than the solid line, which suggests our machine learning models and SDP are reliable. The green dotted line predicted by ensemble learning is lower that the magenta dotted line predicted by SDP. This means that these classifiers find more isotropic states with steerability than their ``master" SDP.

For accurate label datasets, the feature $F'_1$ performs better than feature $F'_2$, in the sense that each machine learning model exhibits higher classify accuracy and robuster generalization capabilities. Therefore, we choose the feature $F'_1$ to predict quantum steering bounds of partially entangled states.

\begin{figure}[h]
	\centering 
	\includegraphics[width=8cm]{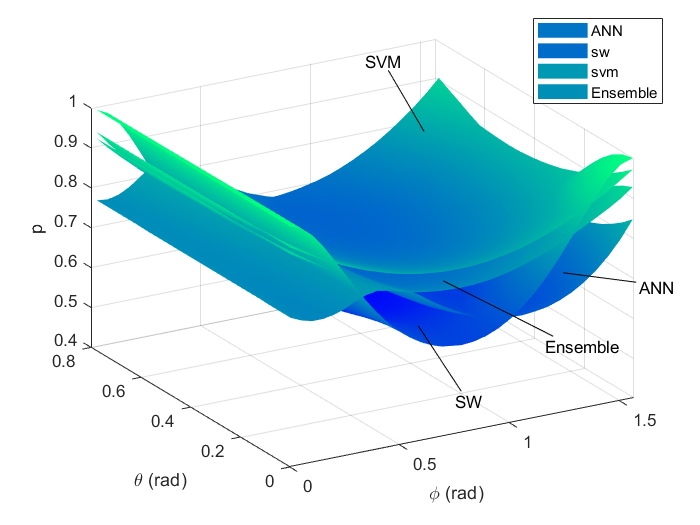}
	\caption{The predictions of steerability for partially entangled state by learned classifiers.}
	\label{Fig.10}
	%\cref{Fig.6}
\end{figure}

Vary $\theta$ and $\phi$, the steerable bounds of partially entangled states can be detected by SVM, ANN, ensemble learning and steer weight defined by SDP. We plot them in Fig.~\ref{Fig.10}. Compared with SVM and ensemble learning, ANN almost always can detect more partially entangled states with steerability. Compared with steering weight, there exist certain regions on which steerable bound of ANN is lower. This showcases ANN's superior predictive capability.

\section{\label{sec:8}Conclusion}

Focus on qutrit-qutrit system, we have constructed the datasets with random states based on the SDP approach, and the datasets with accurate label based on the well-known theory results. For the full information feature $F_1$, ANN model has both high classification accuracy and robust generalisation ability. For these datasets, the steering ellipsoid-like feature $F_2$ and correspond features $F'_1$ and $F'_2$ have been introduced. For the above three features, arbitrary machine learning model in our paper performs well. After analysing the performance of features $F_1$ and $F_2$, we have chosen the feature $F_2$ to predict quantum steering bounds of isotropic states, and have found that the ensemble learning classifiers find more isotropic states with steerability than their ``master" SDP. Similarly, we have chosen the feature $F'_1$ to predict quantum steering bounds of partially entangled states, and have found that there exist certain regions on which steerable bound of ANN is lower than steering weight. This showcases ANN's superior predictive capability. 
Our works offer valuable insights on the detection of quantum steering in high-dimensional systems.

\begin{acknowledgments}
This research was partially supported by
the National Natural Science Foundations of China (Grant
No. 11901317), the China Postdoctoral Science Foundation
(Grant No. 2020M680480), the Fundamental Research Funds
for the Central Universities (Grant No. 2023MS078), and the
Beijing Natural Science Foundation (Grant No. 1232021).

P.W. and Z.Y.L. contributed equally to this work.
\end{acknowledgments}

\appendix

\section{Appendixes}

We list here a few of the tables that appear in the paper.

\begin{table}[H]
	\footnotesize
\caption{\label{tab1}the accuracy of SVM prediction with feature $F_1$}
	%	\doublerulesep 0.1pt \tabcolsep 6pt %space between two columns.
		\begin{ruledtabular}
		\begin{tabular}{cccccc}
			m & 3 & 4 & 5 & 6 & 7 \\
			\hline
			crossrate & 87.90\% & 76.50\% & 85.40\% & 85.10\% & 88.30\% \\
			testrate & 87.20\% & 75.90\% & 85.90\% & 85.70\% & 90.70\%  \\

		\end{tabular}
	\end{ruledtabular}
\end{table}

\begin{table}[H]
	\footnotesize
\caption{\label{tab2}the accuracy of SVM prediction with feature $F_2$}
		%\doublerulesep 0.1pt \tabcolsep 6pt %space between two columns.
		\begin{ruledtabular} 
		\begin{tabular}{cccccc}
			SVM & m=3 & m=4 & m=5 & m=6 & m=7 \\
			\hline
			crossrate & 92.10\% & 87.00\% & 89.40\% & 89.90\% & 88.40\%\\
			testrate & 92.10\% & 87.30\% & 90.90\% & 87.10\% & 88.50\%\\
			Istropicrate & 89.70\% & 91.00\% & 91.10\% & 92.90\% & 92.70\%\\

	\end{tabular}\end{ruledtabular}
\end{table}

\begin{table}[H]
	\footnotesize
\caption{\label{tab4}the accuracy of ANN prediction with feature $F_1$}
	%	\doublerulesep 0.1pt \tabcolsep 6pt
			\begin{ruledtabular}
		\begin{tabular}{cccccc}
			m & 3 & 4 & 5 & 6 & 7 \\
			\hline
			crossrate & 82.10\% & 80.10\% & 80.20\% & 78.10\% & 86.00\% \\
			testrate & 83.70\% & 80.70\% & 80.00\% & 80.30\% & 87.50\%  \\
			Isotropic & 98.50\% & 94.70\% & 98.00\% & 93.10\% & 93.80\%  \\
	\end{tabular}\end{ruledtabular}
\end{table}

\begin{table}[H]
	\footnotesize
\caption{\label{tab5}the accuracy of ANN prediction with feature $F_2$}
	%	\doublerulesep 0.1pt \tabcolsep 6pt
			\begin{ruledtabular}
		\begin{tabular}{cccccc}
			ANN & m=3 & m=4 & m=5 & m=6 & m=7 \\
			\hline
			crossrate & 92.30\% & 87.40\% & 89.50\% & 89.80\% & 89.70\%\\
			testrate & 91.40\% & 87.80\% & 91.50\% & 87.90\% & 88.60\%\\
			Istropicrate & 82.20\% & 87.80\% & 89.50\% & 89.50\% & 88.30\%\\
	\end{tabular}\end{ruledtabular}
\end{table}

\begin{table}[H]
	\centering
	\footnotesize
\caption{\label{tab7}the accuracy of ensemble learning prediction with feature $F_2$}
	%	\doublerulesep 0.1pt \tabcolsep 6pt
			\begin{ruledtabular}
		\begin{tabular}{cccccc}
			Ensemble & m=3 & m=4 & m=5 & m=6 & m=7 \\
			\hline
			crossrate & 92.40\% & 87.30\% & 89.80\% & 89.50\% & 89.80\%\\
			testrate & 91.60\% & 87.90\% & 92.50\% & 87.60\% & 90.40\%\\
			Istropicrate & 91.90\% & 92.90\% & 93.80\% & 93.10\% & 94.90\%\\
	\end{tabular}\end{ruledtabular}
\end{table}

\nocite{*}

\bibliography{reference}% Produces the bibliography via BibTeX.

\end{document}